# Spectrally enhancing near-field radiative transfer between gold gratings by exciting magnetic polariton in nanometric vacuum gaps

Yue Yang and Liping Wang*

*School for Engineering of Matter, Transport, and Energy*
*Arizona State University, Tempe, AZ, USA 85287*

* Corresponding author: liping.wang@asu.edu

**Abstract**

In the present work, we theoretically demonstrate that near-field radiative transport between one dimensional periodic grating microstructures separated by nanometer vacuum gaps can be spectrally enhanced by exciting magnetic polariton. Fluctuational electrodynamics that incorporates scattering matrix theory with rigorous coupled-wave analysis is employed to exactly calculate the near-field radiative flux between two gold gratings. Besides the well-known coupled surface plasmon polaritons, the radiative flux can be also spectrally enhanced due to magnetic polariton, which is excited in the gap between gold ridges. The mechanisms of magnetic polariton in the near-field radiative transport are elucidated in detail, while the unusual enhancement cannot be predicted by either the Derjaguin's or effective medium approximations. The effects of vacuum gap distance and grating geometry parameters between the two gratings are investigated. The findings will open up a new way to control near-field radiative transfer by magnetic polariton with micro/nanostructured metamaterials.

It has been demonstrated during the last decade that, radiative transfer could be significantly enhanced when distance between two objects is smaller than the characteristic thermal wavelength due to photon tunneling or coupling of evanescent waves [1-3]. In particular, near-field radiative flux could far exceed the blackbody limit by the resonant coupling of surface plasmon/phonon polaritons (SPP/SPhP) across the vacuum gap both theoretically and experimentally.[4-6] Recently, excitations of magnetic SPhP [7, 8], hyperbolic modes [9-11], and epsilon-near-pole or epsilon-near-zero modes [12] with different types of metamaterials have also been studied to further improve the near-field radiative flux. Moreover, compared to the case of two plates, the near-field radiative transport between two gratings can be further enhanced due to guided modes [13] and spoof surface plasmon polaritons [14] between two Au gratings, and hyperbolic modes between two doped silicon gratings [15]. Near-field thermal radiation could find many promising applications in energy-harvesting,[1, 16] near-field imaging [17], thermal modulation [18], and thermal switching [15, 19] and rectification [20-22].

Magnetic polaritons (MP) refer to the strong coupling of external electromagnetic waves with the magnetic resonance excited inside the nanostructures. MP artificially realized with metallic micro/nanostructures have been employed to control light propagation and tailor exotic optical and radiative properties in the far field, such as selective solar absorber [23], thermophotovoltaic emitter [24], and switchable or tunable metamaterial [25-27]. Phonon-mediated MP have also been excited in both SiC deep grating and binary grating configurations as well [28]. On the other hand, the MP excitation has been achieved in the $SiO_2$ spacer between two Ag binary gratings [29]. In comparison to SPP/SPhP that has been well studied for tailoring both far- and near-field thermal radiation, magnetic resonance or MP has only been investigated for controlling far-field thermal radiation while its role in near-field radiative transport has yet to be identified.

In this work, we will theoretically investigate the possible effect of MP in near-field radiative transfer between two Au grating microstructures separated by a vacuum gap $d$ below 100 nm. Note that Refs. [13] and [14] also looked into the radiative transfer between two Au gratings but at a large vacuum gap distance of 1 µm, and respectively focused on the heat flux enhancement from guided modes and spoof surface plasmon polaritons.

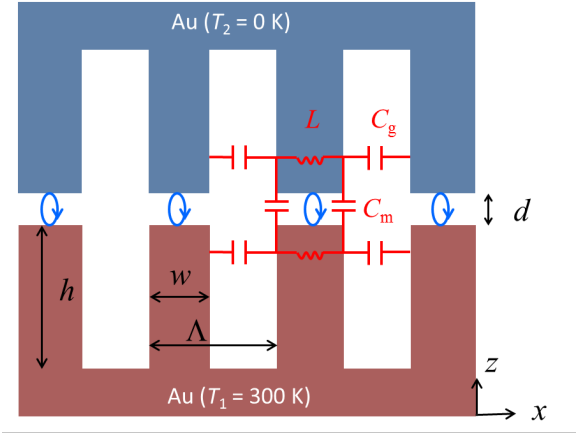

Fig. 1. Schematic of radiative transfer between two symmetric, perfectly aligned Au gratings with parameters like period ($\Lambda$), depth ($h$), and ridge width ($w$). The emitter and receiver temperatures are respectively set as $T_1$ = 300 K and $T_2$ = 0 K. The vacuum gap distance is denoted as $d$. An equivalent LC circuit model and the resulting electrical current loops at excitation of magnetic polariton (MP) are also depicted.

As depicted in Fig. 1, the grating period, depth, and ridge width are kept as $\Lambda = 2$ µm, $h = 1$ µm, and $w = 1$ µm, respectively, which are kept unchanged in the present work unless specified. The grating filling ratio is then $f = w/\Lambda = 0.5$. The temperatures of the emitter and receiver are $T_1$ = 300 K and $T_2$ = 0 K, respectively. Note that an equivalent inductor-capacitor (LC) circuit model, which has been widely used to predict the MP resonance frequency in far field [29], is also shown in Fig. 1 along with the resulting current loop. The question is that whether MP resonance can be excited in the nanometer vacuum gap to enhance near-field radiative transfer. To this end, the scattering formalism [30-32] that is incorporated into fluctuational electrodynamics with rigorous coupled-wave analysis (RCWA) [33, 34] is employed to exactly calculate the near-field radiative flux. The dielectric function of Au is described by a Drude model [35] as $\varepsilon_{Au}(\omega) = 1 - \frac{\omega_p^2}{\omega^2 + i\gamma\omega}$, where $\omega$ is the angular frequency, the plasma frequency is $\omega_p = 1.37 \times 10^{16}$ rad/s, and scattering rate is $\gamma = 7.31 \times 10^{13}$ rad/s at the temperature of 300 K.

Through the exact scattering theory, near-field spectral radiative transfer between two gratings is expressed as [30, 31]

$$q_\omega = \frac{1}{2\pi^3} \Theta(\omega, T_1) \int_0^{\pi/\Lambda} \int_0^\infty \xi(\omega, k_{x0}, k_y) dk_y dk_{x0} \tag{1}$$

where $\Theta(\omega, T_1) = \hbar\omega / (e^{\hbar\omega/k_B T_1} - 1)$ is the Planck oscillator, and $k_{x0}$ and $k_y$ are the incident wavevector components at the grating surface in x and y direction, respectively. Note that $\Theta(\omega, T_2)$ is left out because it equals 0 when $T_2 = 0$ K. The energy transmission coefficient $\xi(\omega, k_{x0}, k_y)$, which considers all the polarization states, is

$$\xi(\omega, k_{x0}, k_y) = \text{tr}(\mathbf{D}\mathbf{W}_1\mathbf{D}^\dagger\mathbf{W}_2) \tag{2a}$$

$$\mathbf{D} = (\mathbf{I} - \mathbf{S}_1\mathbf{S}_2)^{-1} \tag{2b}$$

$$\mathbf{W}_1 = \Sigma_{-1}^{pw} - \mathbf{S}_1\Sigma_{-1}^{pw}\mathbf{S}_1^\dagger + \mathbf{S}_1\Sigma_{-1}^{ew} - \Sigma_{-1}^{ew}\mathbf{S}_1^\dagger \tag{2c}$$

$$\mathbf{W}_2 = \Sigma_1^{pw} - \mathbf{S}_2^\dagger\Sigma_1^{pw}\mathbf{S}_2 + \mathbf{S}_2^\dagger\Sigma_1^{ew} - \Sigma_1^{ew}\mathbf{S}_2 \tag{2d}$$

where $\mathbf{S}_1 = \mathbf{R}_1$ and $\mathbf{S}_2 = e^{ik_{z0}d}\mathbf{R}_2 e^{ik_{z0}d}$. $\mathbf{R}_1$ and $\mathbf{R}_2$ are the reflection operators of the two gratings, which can be obtained through RCWA method [33, 34, 36]. The operators $\Sigma_n^{pw/ew} = \frac{1}{2}k_z^n \Pi^{pw/ew}$, where $\Pi^{pw/ew}$ are the projectors on the propagative and evanescent sectors, were clearly defined in Ref. [31]. Only 1D grating structure with periodicity along $x$ axis is considered here. Note that the wavevector in $x$ direction has been extended from the first Brillouin zone of $k_{x0}$ to infinity through Bloch wave conditions. To ensure the numerical accuracy of the calculation with reasonable computational time, a total of 51 angular frequency values evenly spanned from $3\times10^{14}$ rad/s to $8\times10^{14}$ rad/s was considered, while 21 and 121 data points were used for $k_{x0}$ and $k_y$, respectively, with the upper limit of $k_y$ set as $100\omega/c$ at $d = 100$ nm, for calculating the spectral heat flux at each frequency after double integrations. A total of 361 diffraction orders, which have been checked to be sufficient, were applied to ensure the numerical convergence.

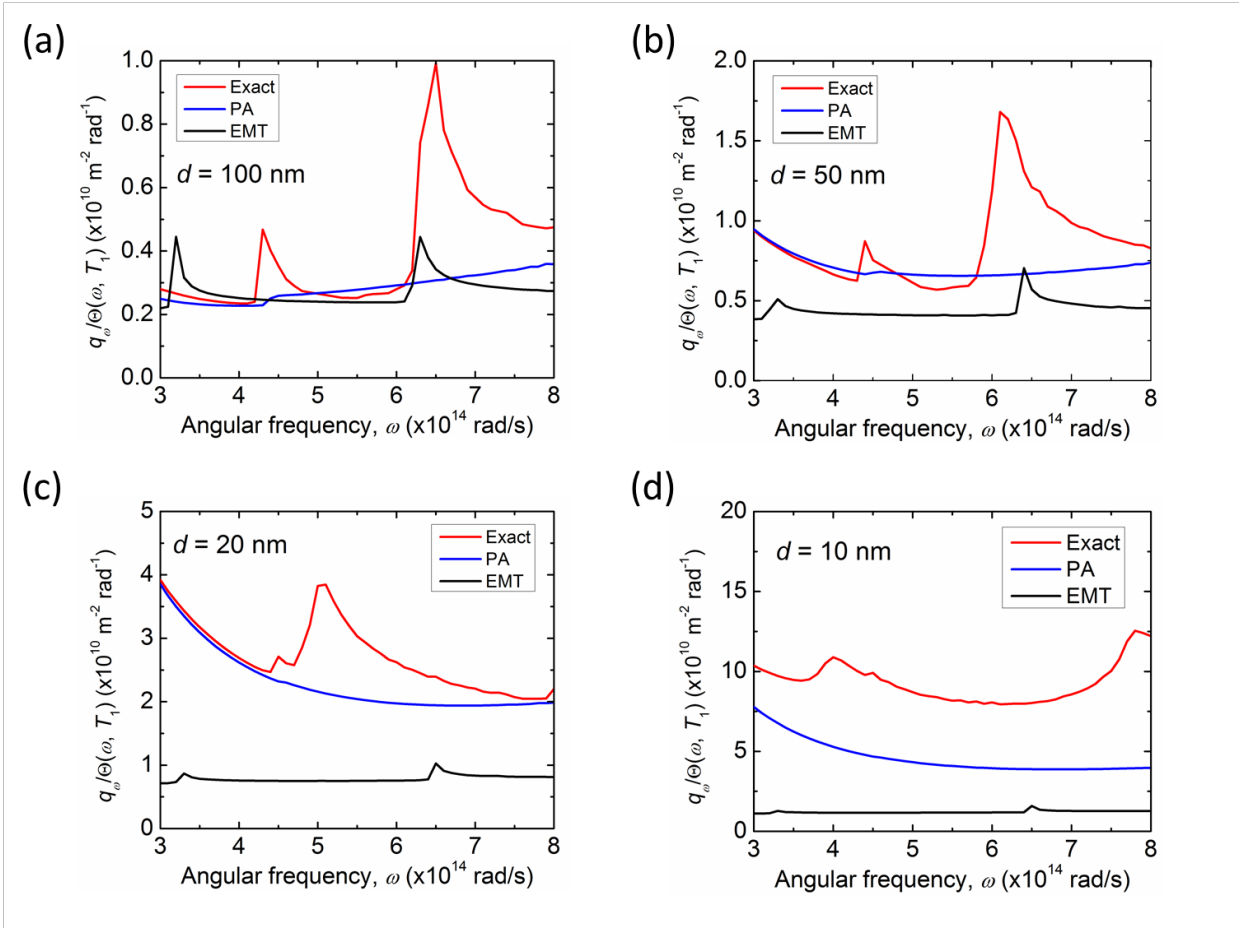

Fig. 2. Normalized spectral heat fluxes between two Au gratings (denoted as exact) at different vacuum gaps of (a) $d = 100$ nm, (b) $d = 50$ nm, (c) $d = 20$ nm, and (d) $d = 10$ nm. The spectral heat fluxes between two Au gratings using Derjaguin's proximity approximation method (denoted as PA), and effective medium theory (denoted as EMT) are also presented.

Figure 2 shows the spectral heat fluxes normalized to the Planck oscillator $\Theta(\omega, T_1)$ between two Au gratings at different vacuum gap distances of $d = 100$ nm, 50 nm, 20 nm and 10 nm. Note that the material and structure effects on heat transfer only impact the normalized spectral heat flux, while the Planck oscillator only considers the effect of temperature. It can be clearly observed from Fig. 2(a) that, at a vacuum gap distance of $d$ = 100 nm, there are two peaks of normalized spectral heat flux at angular frequencies of $4.5\times10^{14}$ rad/s and $6.5\times10^{14}$ rad/s, respectively. When $d$ decreases to 50 nm the larger peak shifts to $\omega = 6.0\times10^{14}$ rad/s, and further to $5\times10^{14}$ rad/s at $d$ = 20 nm. However, when $d$ becomes 10 nm, there are two spectral peaks respectively at the frequencies of $4\times10^{14}$ rad/s and $7.8\times10^{14}$ rad/s, in addition to the small spectral peak around $\omega = 4.5\times10^{14}$ rad/s, whose frequency does not change at all at different vacuum gaps but the peak amplitude increases from 0.5 $m^{-2}\cdot rad^{-2}$ at $d$ = 100 nm to 10 $m^{-2}\cdot rad^{-2}$ at $d$ = 10 nm.

In order to understand the physical mechanisms responsible for the normalized spectral heat flux peaks predicted by the exact calculation, the Derjaguin's proximity approximation (PA) method, which represents a weighted approach for coupled SPP modes with different vacuum gap distances, is first considered. The spectral heat flux between two gratings calculated from the PA method can be weighted by the ones between two plates with different gap distances as

$$q_\omega^{PA} = f \times q_\omega^{plate}(d) + (1-f) \times q_\omega^{plate}(d+2h) \qquad (3)$$

where $q_\omega^{plate}(L)$ means the spectral heat flux between two plates with a gap distance $L$. As inferred by Eq. (3), the PA method only considers the contributions by SPP coupling between planar surfaces at different vacuum gap distances. This indicates that the PA method would be accurate if coupled SPP resonance is the only mechanism that dominates near-field radiative transfer between Au gratings. However, by comparing the normalized spectral heat flux from the PA method to the exact solution in Fig. 2, the PA method turns out to be accurate with good agreement with the exact solution except for the angular frequencies where spectral heat flux peaks exist for $d$ from 100 nm to 20 nm. At $d$ = 10 nm, the PA method fails to predict the exact values by significant discrepancies within the entire spectrum of interests. Apparently, the PA method or the coupled SPP mode between planar surfaces cannot explain the spectral heat flux peaks that exist between gold gratings.

Furthermore, the effective medium theory (EMT), which considers the grating layer as a homogeneous uniaxial medium, is also examined on whether or not to be responsible for the spectral enhancement. The effective dielectric functions of the homogenized thin film for the grating region are expressed as

$$\varepsilon_\text{O} = (1-f) + \varepsilon_\text{Au} f \tag{4a}$$

$$\varepsilon_\text{E} = \frac{\varepsilon_\text{Au}}{(1-f)\varepsilon_\text{Au} + f} \tag{4b}$$

where the subscript "O" and "E" denote the ordinary and extraordinary component of dielectric function, respectively. As the optical axis of considered grating structure is along x axis, it gives $\varepsilon_{xx} = \varepsilon_\text{E}$, and $\varepsilon_{yy} = \varepsilon_{zz} = \varepsilon_\text{O}$. Due to the anisotropy between x and y directions, cross polarizations and thin-film optics have to be considered when calculating the radiative transfer between uniaxial films [15]. As shown in Fig. 2, the normalized spectral heat fluxes predicted by EMT exhibit two spectral peaks around $\omega = 3.3 \times 10^{14}$ rad/s and $6.5 \times 10^{14}$ rad/s, whose frequencies and magnitudes little change with $d$ from 100 nm to 10 nm, indicating that the vacuum gap distance has negligible effect on both spectral peaks. In fact, those two peaks predicted by EMT are actually due to the coupled SPP modes or called bulk polariton within the grating layer approximated as an effective homogenous uniaxial film. However, both spectral peaks and the spectral heat flux spectra from the EMT prediction cannot match the exact calculation. More importantly, the red-shift behavior of the larger spectral peak with smaller vacuum gaps from the exact solution cannot be captured by EMT at all. After all, EMT is inherently a homogenization approach which cannot take into account the local resonance modes like coupled SPP or MP that could possibly occur within the vacuum gap [15, 37, 38]. Therefore, the effective medium approximation cannot explain the unusual radiative transfer between gold gratings across ultrasmall vacuum gaps, while physical mechanisms other than coupled SPP and EMT have to be identified and understood here.

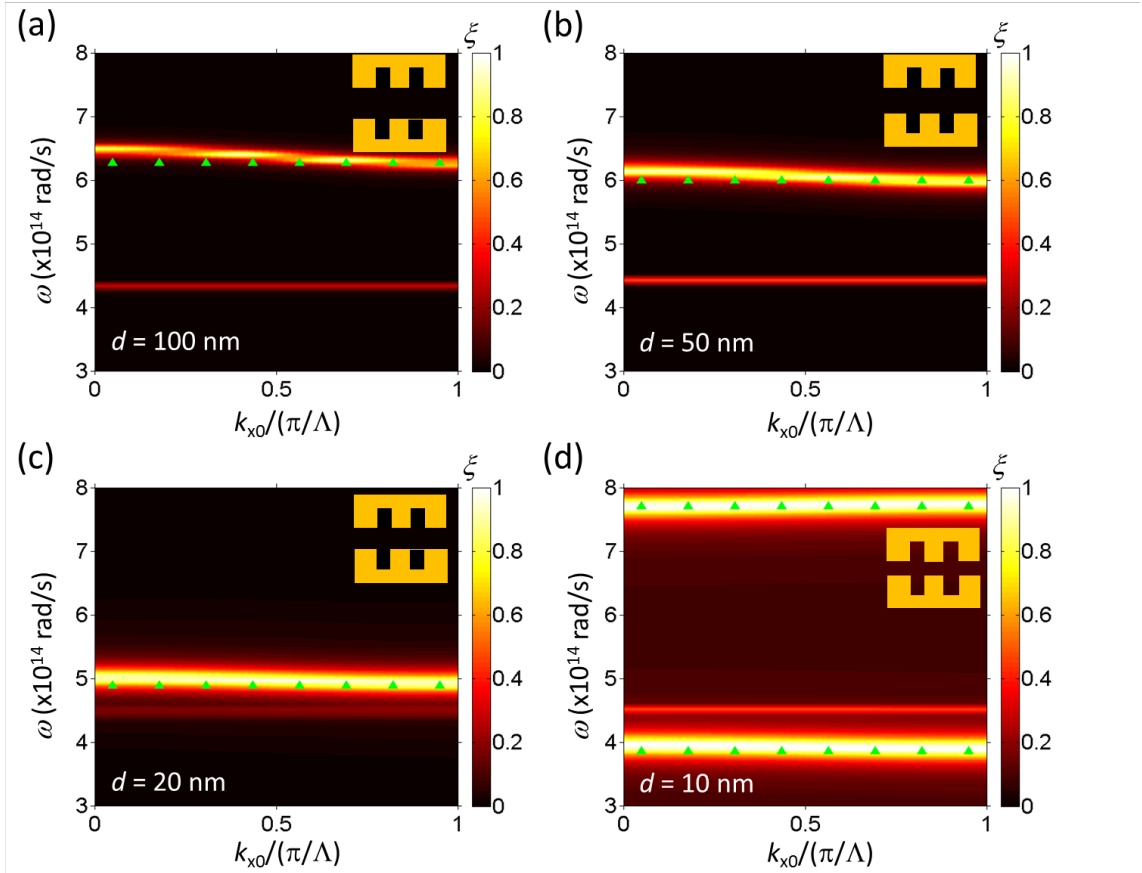

Fig. 3. Contour plots of energy transmission coefficient ($\xi$) between two Au gratings at vacuum gaps of (a) $d = 100$ nm, (b) $d = 50$ nm, (c) $d = 20$ nm, and (d) $d = 10$ nm. The base geometric parameters of Au gratings are $\Lambda = 2$ μm, $w = 1$ μm, and $h = 1$ μm. Note that $k_y = 0$ is assumed and $k_{x0}$ is normalized to the first Brillouin zone. The LC circuit model prediction of MP resonance conditions is shown as green triangles.

To gain a better idea on the radiative transfer between gold gratings, the contour plots of transmission coefficient in the $\omega$ - $k_{x0}$ domain under $k_y = 0$ from the exact method are presented in Fig. 3 at corresponding vacuum gap distances. Multiple bright horizontal bands, which are independent of $k_{x0}$ and indicate the enhanced near-field radiative transfer channels, can be clearly observed. Among different vacuum gap distances, the one at $\omega = 4.5\times10^{14}$ rad/s barely shifts with different d values, which is corresponding to the smaller spectral heat flux peak at the same frequency observed in Fig. 2. As intensively discussed in Ref. [13], this is associated with the guided mode, whose resonance condition strongly depends on the cavity depth, i.e. $H = 2h+d$. Note that the grating depth is $h = 1$ μm, which is much larger than the sub-100-nm vacuum gap distances considered here. Therefore, it can be understood that, the guide mode would not shift when $d$ varies from 100 nm to 10 nm as $H ? d$. However, the same theory of guided modes cannot explain the brighter and broader resonance mode around $\omega = 6.5\times10^{14}$ rad/s at $d = 100$ nm, which clearly shifts to lower frequencies with smaller $d$. As the coupled SPP, effective medium and guided modes cannot explain this particular unusual spectral enhancement between gold grating, could it be associated with possible excitation of magnetic resonance or MP?

In order to verify our hypothesis of MP resonance, an equivalent LC circuit is employed to analytically predict the resonance conditions of MP between two Au gratings in near field [29]. Note that the LC model, based on the resonant charge distributions, has been successfully employed to verify the physical mechanisms of MP modes in metal-insulator-metal (MIM) nanostructures in selective control of far-field thermal radiation [26-29]. After all, the nanometer vacuum gap between the two Au grating here forms similar MIM configurations. Here, the inductance of Au grating can be expressed as $L = L_m + L_e$, where the first term $L_m = 0.5\mu_0 wd$ accounts for the mutual inductance of two parallel plates with width $w$ separated by a distance $d$, and the kinetic inductance $L_e = \omega/(\varepsilon_0 \omega_p^2 \delta)$ considers the contribution of drifting electrons. Note that $\mu_0$ and $\varepsilon_0$ are the permeability and permittivity of vacuum, while $\delta = \lambda/2\pi\kappa$ is the field penetration depth with $\kappa$ being the extinction coefficient of Au. On the other hand, the parallel-plate capacitance between the upper and lower Au ridges can be expressed as $C_m = c_1\varepsilon_0 w/d$, where $c_1$ = 0.22 is the correction factor considering non-uniform charge distribution.[29] The capacitance between left and right Au ridges is denoted as $C_g = \varepsilon_0 h/(\Lambda - w)$. Thus, the resonance frequency for the fundamental MP mode can be obtained when the total circuit impedance reaches zero:

$$\omega_{MP1} = 1/\sqrt{(L_m + L_e)(C_m + C_g)} \tag{5}$$

With the base grating geometries as $\Lambda$ = 2 µm, $w$ = 1 µm, $h$ = 1 µm, the MP1 resonance frequencies between two Au gratings are predicted to be 6.4×10$^{14}$ rad/s, 6.0×10$^{14}$ rad/s, 4.9×10$^{14}$ rad/s, and 3.8×10$^{14}$ rad/s respectively for $d$ = 100 nm, 50 nm, 20 nm, and 10 nm, which match surprisingly well with the unusual spectral enhancement mode predicted by the exact solution as shown in Fig. 3. Note that the independence of the MP resonance condition on the $k_{x0}$ has been thoroughly discussed and well understood previously [28, 29]. At $d$ = 10 nm, the contour shows an additional resonance mode around $\omega$ = 7.7×10$^{14}$ rad/s, which is actually the second harmonic order of MP resonance with doubled resonance frequency from MP1. The unanimous agreements between the exact solution and the analytical LC model prediction at different vacuum gaps clearly verify the physical mechanism of MP excitation in spectrally enhancing near-field radiative transfer between Au grating structures.

To further confirm and understand the behaviors of MP resonance in near-field radiative transport across nanometer vacuum gaps, the grating geometric effect on near-field radiative transfer between Au gratings is investigated in terms of transmission coefficient at the gap distance $d$ = 20 nm. Figure 4(a) and 4(b) present the effect of grating depth respectively with $h$ = 0.5 µm and 1.5 µm, while other geometric parameters are kept at the base values. In comparison with the case of $h$ = 1 µm in Fig. 3(c), the strong and broad MP resonance mode around $\omega$ = 5×10$^{14}$ rad/s slightly shifts toward lower frequencies, which is in good agreement with the LC model prediction. Note that, the grating depth $h$ only affects the capacitance $C_g$, which is less than $C_m/10$ and thereby negligible with given parameters. Therefore, $h$ has little effect on the MP resonance condition. On the other hand, the guided mode, which strongly depends on the cavity depth, shifts from $\omega = 4.5\times10^{14}$ rad/s at $h = 1$ μm to $6.2\times10^{14}$ rad/s when grating depth becomes 1.5 μm.

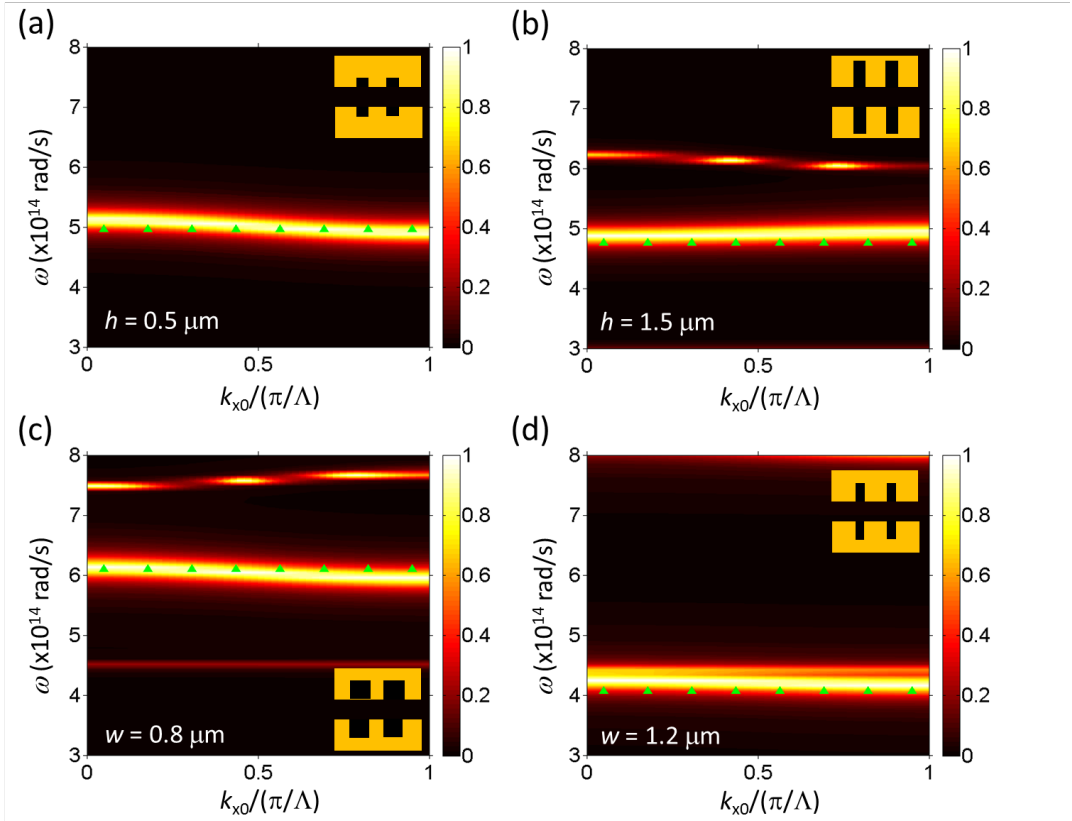

Fig. 4. Contour plots of energy transmission coefficient ($\xi$) between two Au gratings at $d = 20$ nm with different geometries: (a) $h = 0.5$ μm, (b) $h = 1.5$ μm, (c) $w = 0.8$ μm, and (d) $w = 1.2$ μm, while the rest of parameters are kept at the base values: $\Lambda = 2$ μm, $w = 1$ μm, and $h = 1$ μm. Note that $k_y = 0$ is assumed and $k_{x0}$ is normalized to the first Brillouin zone. The LC circuit model prediction of MP resonance conditions is also shown as green triangles.

Finally, the effect of grating width on the near-field radiative transfer spectrally enhanced by MP resonance is studied similarly in terms of transmission coefficient. By comparing Figs. 4(c) and 4(d) with 3(c), when the grating ridges becomes wider from 0.8 to 1.2 μm, both the exact solution and the LC model consistently show that the MP resonance mode changes from $\omega$ = $6.1\times10^{14}$ rad/s to $4.2\times10^{14}$ rad/s. From the perspective of charge distribution, grating width ($w$) is linear to $L_m$, $L_e$, and $C_m$. With negligible $C_g$ at $d$ = 20 nm, the MP resonance frequency $\omega_{MP1} \approx 1/\sqrt{(L_m + L_e)C_m}$ is essentially inversely proportional to $w$. On the contrary, the weaker guided mode around $\omega$ = $4.5\times10^{14}$ rad/s does not change with different grating width, whose resonance frequency is only a strong function of cavity depth H or grating depth $h$ [13].

In summary, we have theoretically demonstrated unusual spectral radiative flux enhancement between two gold gratings separated by sub-100-nm vacuum gaps, which neither coupled SPP, effective medium, nor guided mode could explain. The physical mechanisms have been identified and elaborated for the first time to be the excitation of magnetic polariton within the vacuum gap for spectrally enhancing the near-field radiative transport. The vacuum gap and geometric dependences of the MP modes in tuning near-field radiative transfer are consistent between the exact solutions and the predictions from an LC circuit model, which unarguably verifies the mechanisms of MP for controlling near-field thermal radiation. The fundamental understanding gained here will open up a new way to spectrally tailor near-field radiative transfer with metamaterials for thermal management and energy harvesting applications.

**ACKNOWLEDGMENT**

This work was supported by the National Science Foundation under CBET-1454698.

**REFERENCES**

[1]   S. Basu, Z. M. Zhang, and C. J. Fu, Int. J. Energy Res. **33**, 1203 (2009).
[2]   D. G. Cahill, P. V. Braun, G. Chen, D. R. Clarke, S. H. Fan, K. E. Goodson, P. Keblinski, W. P. King, G. D. Mahan, and A. Majumdar, Appl. Phys. Rev. **1**, 011305 (2014).
[3]   D. G. Cahill, W. K. Ford, K. E. Goodson, G. D. Mahan, A. Majumdar, H. J. Maris, R. Merlin, and S. R. Phillpot, J. Appl. Phys. **93**, 793 (2003).

[4] C. Fu and Z. Zhang, Int. J. heat Mass Tran. **49**, 1703 (2006).

[5] S. Shen, A. Mavrokefalos, P. Sambegoro, and G. Chen, Appl. Phys. Lett. **100**, 233114 (2012).

[6] S. Shen, A. Narayanaswamy, and G. Chen, Nano Lett. **9**, 2909 (2009).

[7] S. Basu and M. Francoeur, Appl. Phys. Lett. **99**, 143107 (2011).

[8] S. J. Petersen, S. Basu, and M. Francoeur, Photonics Nanostruct. **11**, 167 (2013).

[9] S.-A. Biehs, M. Tschikin, and P. Ben-Abdallah, Phys. Rev. Lett. **109**, 104301 (2012).

[10] C. Cortes, W. Newman, S. Molesky, and Z. Jacob, J. Opt. **14**, 063001 (2012).

[11] Y. Guo, C. L. Cortes, S. Molesky, and Z. Jacob, Appl. Phys. Lett. **101**, 131106 (2012).

[12] S. Molesky, C. J. Dewalt, and Z. Jacob, Opt. Express **21**, A96 (2013).

[13] R. Guérout, J. Lussange, F.S.S. Rosa, J.-P. Hugonin, D.A.R. Dalvit, J.-J. Greffet, A. Lambrecht and S. Reynaud, J. Phy. Conf. Ser. **395**, 012154 (2012).

[14] J. Dai, S.A. Dyakov and M. Yan, Phys. Rev. B **92**, 035419 (2015).

[15] X.L. Liu, B. Zhao and Z.M. Zhang, Phys. Rev. A **91**, 062510 (2015).

[16] K. Park, S. Basu, W. P. King, and Z. M. Zhang, J. Quant. Spectrosc. RA. **109**, 305 (2008).

[17] K. Hoshino, A. Gopal, M. S. Glaz, D. A. Vanden Bout, and X. Zhang, Appl. Phys. Lett. **101**, 043118 (2012).

[18] P. Ben-Abdallah and S.-A. Biehs, Phys. Rev. Lett. **112**, 044301 (2014).

[19] W. Gu, G.H. Tang and W.Q. Tao, Int. J. heat Mass Tran. **82**, 429 (2015).

[20] C. R. Otey, W. T. Lau, and S. Fan, Phys. Rev. Lett. **104**, 154301 (2010).

[21] L. P. Wang and Z. M. Zhang, Nanosc. Microsc. Therm. **17**, 337 (2013).

[22] Y. Yang, S. Basu, and L. Wang, Appl. Phys. Lett. **103**, 163101 (2013).

[23] H. Wang and L. Wang, Opt. Express **21**, A1078 (2013).

[24] B. Zhao, L. Wang, Y. Shuai, and Z. M. Zhang, Int. J. Heat Mass Tran. **67**, 637 (2013).

[25] H. Wang, Y. Yang, and L. Wang, Appl. Phys. Lett. **105**, 071907 (2014).

[26] H. Wang, Y. Yang and L.P. Wang, J. Appl. Phy. **116**, 123503 (2014).

[27] H. Wang, Y. Yang and L.P. Wang, J. Opt. **17**, 045104 (2015).

[28] L. P. Wang and Z. M. Zhang, Opt. Express **19**, A126 (2011).

[29] L.P. Wang and Z.M. Zhang, JOSA B **27**, 2595 (2010).

[30] A. Lambrecht and V. N. Marachevsky, Phys. Rev. Lett. **101**, 160403 (2008).

[31] J. Lussange, R. Guérout, F. S. S. Rosa, J. J. Greffet, A. Lambrecht, and S. Reynaud, Phys. Rev. B **86**, 085432 (2012).

[32] J. Lussange, R. Guérout, and A. Lambrecht, Phys. Rev. A **86**, 062502 (2012).

[33] L. Li, J. Opt. Soc. Am. A. **13**, 1870 (1996).

[34] M. G. Moharam, E. B. Grann, D. A. Pommet, and T. K. Gaylord, J. Opt. Soc. Am. A **12**, 1068 (1995).

[35] L.P. Wang, S. Basu and Z.M. Zhang, J. Heat Tran. **134**, 072701 (2012).

[36] R. Guérout, J. Lussange, H.B. Chan, A. Lambrecht and S. Reynaud, Phys. Rev. A **87**, 052514 (2013)

[37] H. Chalabi, E. Hasman and M.L. Brongersma, Phys. Rev. B **91**, 014302 (2015).

[38] X.L. Liu, T.J. Bright and Z.M. Zhang, J. Heat Tran. **136**, 092703 (2014).